\documentclass[preprint,18pt,Times]{revtex4-1}
\usepackage{amsmath,natbib,graphicx,amssymb,graphics,amsmath,mathrsfs,CJK,color}
\usepackage{multirow,fancyhdr,color,bm,tabularx,psfrag,dcolumn}
\usepackage[colorlinks=true,linkcolor=blue,urlcolor=blue,citecolor=blue]{hyperref}
\usepackage[left]{lineno}%[switch]
\begin{document}
%\begin{CJK*}{GBK}{song}
\preprint{\emph{ Submit to  Chinese Physics B }}
%%****************************************************************
\title{ Influence of the coupled-dipoles on photosynthetic performance in a photosynthetic quantum heat engine }%\LARGE\boldmath\bf
%\thanks{}
\author{Ling-Fang Li  }
\affiliation{Department of Physics, Faculty of Science, Kunming University of Science and Technology, Kunming, 650500, PR China}

\author{Shun-Cai Zhao }
\email[Corresponding author: ]{zhaosc@kmust.edu.cn.}
\affiliation{Department of Physics, Faculty of Science, Kunming University of Science and Technology, Kunming, 650500, PR China}

%%****************************************************************
%\author{ }
%\affiliation{Department of Physics, Faculty of Science, Kunming University of Science and Technology, Kunming, 650500, PR China}
\date{\today}

%%****************************************************************
%\linenumbers
\begin{abstract}
Recent evidences suggest that the multi charge-separation pathways can contribute to the photosynthetic performance. In this work, the influence of coupled-dipoles on the photosynthetic performance was investigated in a two-charge separation pathways quantum heat engine (QHE) model. And the population dynamics of the two coupled sites, j-V characteristics and power involving this photosynthetic QHE model were evaluated for the photosynthetic performance. The results illustrate that the photosynthetic performance can be greatly enhanced but quantum interference was deactivated by the coupled-dipoles between the two-charge separation pathways. However, the photosynthetic performance can also be promoted by the deactivated quantum interference owing to the coupled-dipoles. It is a novel role of the coupled-dipoles in the energy transport process of biological photosynthetic and some artificial strategies may be motivated by this photosynthetic QHE model in future.
\begin{description}
\item[PACS numbers]{ 42.50.Gy, 42.50.-p, 32.80.Qk }
\item[Keywords]{ photosynthetic performance; coupled-dipole; photosynthetic heat engine }
\end{description}
\end{abstract}
\maketitle
\section{Introduction}

Charge separation is an essential processes in the conversion of solar energy into chemical energy in photosynthesis\cite{EngelEvidence}. After absorption of a photon which creates an electronically excited state in a pigment molecule, the excited state is transformed into a short-lived charge-separated state until it arrives at a reaction center (RC)\cite{AkihitoTheoretical,ZhuModified,Caram2012Signatures,RebentrostEnvironment,KilloranEnhancing} in the pigment-protein complex and results in a stable charge-separated state, which ultimately powers the photosynthetic organism\cite{MohseniEnvironment}. The photon-to-charge conversion efficiency in above process is considered to be close to \(100\%\) under certain conditions\cite{Est2002Blankenship,GilmoreQuantum,CarusoHighly,ChenVibration}. This eye-catching result sparks the researchers to explore its physics behind the energy conversion within photosynthesis\cite{Collini2010,Romero2014}. People firmly believe that understanding its underlying mechanism involving photosynthesis can help us in designing novel artificial nano-devices for efficient quantum transport and some optimized solar cells\cite{Zhao2019,Fingerhu2010Phys,2020Inhibited,Blankenship2011Scie,zhao2020EPJP}. Recently, some theoretical models\cite{FlemingTheoretical,PlenioVibrations,2018Enhancing} show that the quantum mechanism behavior contributes beneficially to the high efficiency of biological process.

Simulating the photosynthetic RC as a biological QHE, Dorfman et al.\cite{DorfmanPhotosynthetic} suggested that photocurrent can be improved by noise-induced coherence in the photosynthetic RC. Creatore et al.\cite{CreatoreAn} have proposed a model for photosynthetic RC in which quantum mechanical effects were investigated. The results show that photo-to-charge conversion can be boosted via quantum interference caused by dipole-dipole interactions\cite{2016Angle} between molecular excited states. And compared with classical photocell, this effect can increase the current and power output to 35\%. In a multiple charge-separation pathways scheme\cite{Qin2016A}, the results demonstrated that a multi-pathway biological QHE was a better choice to the charge separation then help to improve the quantum current and power yield. Recently, long-lived quantum coherence in the pigment-protein complexes was observed and extensively investigated in some experiments\cite{PanitchayangkoonLong,ColliniCoherently,EngelEvidence,Huo2010Iterative,ChristenssonOrigin}.

Although electron transfer in the photosynthetic RC has been thoroughly investigated in one or several charge-separation pathways, and it is believed that two main pathways make significant contribution to the current and power in photosynthetic process\cite{CARDONA201226}. However, works on the influence of the coupled-dipoles between two different charge separation pathways are rarely found. Therefore, we consider the coupled-dipoles between two different charge separation pathways involving the photosynthetic performance in a double-pathway\cite{Arp2015Natural} photosynthetic QHE model, which is different from the work discussed the system-bath couplings effect on the exciton-transfer processes in the Photosystem II reaction center via the polaron master-equation approach\cite{M2017Effects} in a QHE model. We no longer care about the impact of multi-charge-separation pathways\cite{CARDONA201226,Yi2015A}, but turn our attention to the impact of the coupled-dipoles on the photosynthetic performance. Focusing on this issue whether the coupled-dipoles is more beneficial to enhance the photosynthetic performances in this two-pathways biological QHE model, which can be use to mimic the process in the RC of the light-harvesting complex Phycocyanin-645 (PC645) with a pair of strongly coupled sites (called DBV C and DBV D ) \cite{Collini2010,BrumerPhysical}.

\section{Model and equation}

\begin{figure}
\center
\includegraphics[width=0.45\columnwidth]{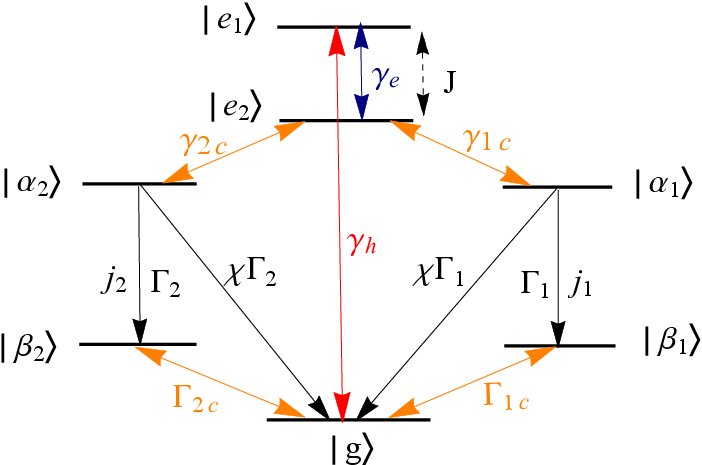 }
\caption{(Color online) Energy-level framework of photosynthetic QHE for the photosynthetic RC with two load-transitions \(|\alpha_{i}\rangle\)\(\rightarrow\)\(|\beta_{i}\rangle_{i=1,2}\). The electronic transition from the ground state \(|g\rangle\) to two coupled dipoles \(|e_{1}\rangle\) ,\(|e_{2}\rangle\) is induced by the high temperature photon bath. The low temperature phonon bath drives charge transfer from the level \(|e_{2}\rangle\) \(\rightarrow\) \(|\alpha_{i}\rangle_{i=1,2}\), and \(|\beta_{i}\rangle_{i=1,2}\) \(\rightarrow\)\(|g\rangle\) with termination of  the electronic circulation. }
\label{Fig.1}
\end{figure}

Here, we investigate the influence of the coupled-dipoles in a two-pathway QHE model illustrated in Fig.\ref{Fig.1}, which is abstracted from the Photosystem II reaction center (PSII-RC) consisting of six core pigment molecules. The typical PSII-RC found in purple bacteria and in oxygen-evolving organisms (cyanobacteria, algae, and higher plants) contains six pigment molecules coupled by the dipole-dipole interactions\cite{NovoderezhkinMixing,ElisabetTwo,Novoderezhkin2015Multiple}, and the six pigments are arranged in two symmetric branches of protein matrix in the center of the complex. Four chlorophylls of them (special pair PD\(_{1}\), PD\(_{2}\) and accessory ChlD\(_{1}\), ChlD\(_{2}\)) and two pheophytins (PheD\(_{1}\), PheD\(_{2}\)), are parallel distributed in these two branches of protein matrix. The pair of chlorophylls, PD\(_{1}\), PD\(_{2}\) located at the center of the PSII RC act as the primary electron donors, forming two exciton states denoted as \(|e_{1,2}\rangle\) in Fig.\ref{Fig.1}. Two pheophytin pigments, PheD\(_{1}\) and PheD\(_{2}\) couple to the rest of the molecules and Phe\(D_{1}\) plays the part of electron acceptor\cite{NovoderezhkinMixing,ElisabetTwo,Novoderezhkin2015Multiple} in the charge-separation process of  the PSII RC.

As shown in Fig.\ref{Fig.1}, after the absorption of a solar photon, the excited electron is promoted from \(|g\rangle\) to \(|e_{2}\rangle\) and/or \(|e_{1}\rangle\)  with the transition rate \(\gamma_{h}\), and the excited electron may transit to \(|e_{1}\rangle\) state at the rate \(\gamma_{e}\). Then the excited electron is transferred to the acceptors by emission of a phonon via the two pathways: \(|e_{1}\rangle\)(\(|e_{2}\rangle\))\(\rightarrow\) \(|\alpha_{1}\rangle\) and \(|e_{1}\rangle\)(\(|e_{2}\rangle\))\(\rightarrow\) \(|\alpha_{2}\rangle\) at the emissions rates \(\gamma_{ic (i=1,2)}\) with \(|\alpha_{1}\rangle\) and \(|\alpha_{2}\rangle\) being the ion-pair states in these two pathways. Furthermore, the positive and negative charges are spatially separated by releasing the excited electron to the plastoquinone molecule and leaving a hole in the dimer. Eventually the electron is transferred to the final electron acceptor PheD\(_{1}\) from which the electron is released to perform work\cite{CreatoreAn} with a rate \(|\Gamma_{i}\rangle_{(i=1,2)}\), denoted by \(|\alpha_{1}\rangle\) \(\rightarrow\) \(|\beta_{1}\rangle\) (Path 1) and \(|\alpha_{2}\rangle\) \(\rightarrow\) \(|\beta_{2}\rangle\) (Path 2) in Fig.\ref{Fig.1}. Finally, the electron returns to the primary electron donor via \(|\beta_{1}\rangle\)(\(|\beta_{2}\rangle\)) \(\rightarrow\) \(|g\rangle\). Similarly, the acceptor-to-donor charge recombination for two pathways, described by \(|\alpha_{1}\rangle\)(\(|\alpha_{2}\rangle\)) \(\rightarrow\) \(|g\rangle\) with rate  \(\chi \Gamma_{i=1,2}\), brings the system back to the ground state \(|g\rangle\) but does not produce current, where \(\chi\) is a dimensionless fraction\cite{Zhao2020} describing the radiative recombination rate of the two pathways. For the physical prospective, we assume the load transition is unidirectional, which means that only the photosynthetic heat engine can transfer energy to the load, no other ways around. Therefore, the total current is given by $j= e \sum\limits_{i=1}^{2} \Gamma_{i}\rho_{\alpha_{i}\alpha_{i}}$  with \(e\) being the elementary electron charge.

For the purpose of clarifying the influence caused by the coupled-dipoles, we will evaluate the photosynthetic performances via the current and power involving this photosynthetic QHE numerically. Therefore, we considered the dipole-dipole coupling depicted by two excited states \(|e_{1/2}\rangle\)=\(\frac{1}{\sqrt{2}}(|D_{1}\rangle\pm|D_{2}\rangle)\)\cite{Hong2014Long} between the two different charge separation pathways with their eigen-energies \(E_{e1/e2}\)=\(E_{D_{1}/D_{2}}\pm J \), where J depicts the coupling robustness illustrated in Fig.\ref{Fig.1}. The state \(|D_{i=1,2}\rangle\) describes chlorophylls PD\(_{i}\),  the electron donors. With this knowledge, the Hamiltonian of this QHE model is consisted of three parts: the electronic Hamiltonian is read as,

\begin{align}
\hat{H}_{e}=&E_{g}|g\rangle\langle g|+\sum^{2}_{i=1}(E_{\alpha_{i}}|\alpha_{i}\rangle\langle \alpha_{i}|+E_{\beta_{i}}|\beta_{i}\rangle\langle \beta_{i}|+E_{D_{i}}|D_{i}\rangle\langle D_{i}|)+ J(|D_{1}\rangle\langle D_{2}|+|D_{2}\rangle\langle D_{1}|),
\end{align}

\noindent A full microscopic mode Hamiltonian is given as follows,

\begin{align}
\hat{H}_{0}=\sum_{k}\hbar\omega_{k}\hat{a}^{\dag}_{k}\hat{a}_{k},
\end{align}

\noindent Considering above modes and ambient reservoirs linearly coupled to the electrical system\cite{Zhao2020,ZHAO2020106329}, the interaction Hamiltonian can be written in rotating-wave approximation as follows,

\begin{align}
\hat{H}_{v}=\hat{V}_{H}+\hat{V}_{1c}+\hat{V}_{2c}+\widehat{V}_{3c},
\end{align}

\noindent where the items in above expression are given in the following forms,

\begin{align}
&\hat{V}_{H}=\sum_{k}\hbar(\varepsilon_{hk}\hat{\sigma}_{g1}\otimes\hat a_{hk}^{\dag}+\varepsilon_{hk}^{\ast}\hat{\sigma}_{g1}^{\dag}\otimes\hat a_{hk})\nonumber, \\
&\hat{V}_{1c}=\sum^{2}_{i=1}\sum_{k}\hbar(\varepsilon_{1ck}\hat{\sigma}_{i2}\otimes\hat a_{1ck}^{\dag}+\varepsilon_{1ck}^{\ast}\hat{\sigma}_{i2}^{\dag}\otimes\hat a_{1ck}), \\
&\hat{V}_{2c}=\sum^{2}_{i=1}\sum_{k}\hbar(\varepsilon_{2ck}\hat{\sigma}_{gi}\otimes\hat a_{2ck}^{\dag}+\varepsilon_{2ck}^{\ast}\hat{\sigma}_{gi}^{\dag}\otimes\hat a_{2ck})\nonumber, \\
&\hat{V}_{3c}=\sum_{k}\hbar(\varepsilon_{3ck}\hat{\sigma}_{21}\otimes\hat  a_{3ck}^{\dag}+\varepsilon_{3ck}^{\ast}\hat{\sigma}_{21}^{\dag}\otimes \hat{a}_{3ck})\nonumber,
\end{align}

\noindent where $\varepsilon_{ik}(i=h,1c,2c,3c)$ is the corresponding couple strength of pigment i to the \emph{k}th mode of reservoir, $\hat a_{ik}^{\dag}(\hat a_{ik})(i=h,1c,2c,3c)$ are the creation(annihilation) operator of \emph{k}th reservoir mode, the system operators are defined as $\hat{\sigma}_{g1}=|g\rangle\langle e_{1}|,\hat{\sigma}_{i2}=|\alpha_{i}\rangle\langle e_{2}|(i=1,2),\hat{\sigma}_{gi}=|g\rangle\langle \beta_{i}|_{(i=1,2)}$, and $\hat{\sigma}_{21}=|e_{2}\rangle\langle e_{1}|$.

The total system Hamiltonian, \(\hat{H}_{T}\)=\(\hat{H}_{e} + \hat{H}_{0} + \hat{H}_{v}\), where the electronic Hamiltonian \(\hat{H}_{e}\) describes the unitary evolution of the electron transfer via the Lindblad-type master equation,

\begin{equation}
\frac{d\hat{\rho}}{dt}=-i[\hat{H}_{e},\hat{\rho}]+\mathscr{L}_{H}\hat{\rho}+\mathscr{L}_{1c}\hat{\rho}+\mathscr{L}_{2c}\hat{\rho}+\mathscr{L}_{3c}\hat{\rho}
+\mathscr{L}_{\Gamma}\hat{\rho}+\mathscr{L}_{\lambda\Gamma}\hat{\rho},
\end{equation}

\noindent The Lindblad-type superoperators in Eq.(5) are listed below,

\begin{eqnarray}
&\mathscr{L}_{H}\hat{\rho}=&\frac{\gamma_{h}}{2}[(n_{h}+1)(2\hat{\sigma}_{g1}\hat{\rho}\hat{\sigma}_{g1}^{\dag}-\hat{\sigma}_{g1}^{\dag}\hat{\sigma}_{g1}
                           \hat{\rho}-\hat{\rho}\hat{\sigma}_{g1}^{\dag}\hat{\sigma}_{g1})\nonumber\\
&&+n_{h}(2\hat{\sigma}_{g1}^{\dag}\hat{\rho}\hat{\sigma}_{g1}-\hat{\sigma}_{g1}\hat{\sigma}_{g1}^{\dag}\hat{\rho}-\hat{\rho}\hat{\sigma}_{g1}\hat{\sigma}_{g1}^{\dag})],\\
&\mathscr{L}_{1c}\hat{\rho}=&\sum^{2}_{i,j=1}\frac{\gamma_{ijc}}{2}[(n_{1c}+1)(\hat{\sigma}_{j2}\hat{\rho}\hat{\sigma}_{i2}^{\dag}
    +\hat{\sigma}_{i2}\hat{\rho}\hat{\sigma}_{j2}^{\dag}-\hat{\sigma}_{j2}^{\dag}\hat{\sigma}_{i2}\hat{\rho}-\hat{\rho}\hat{\sigma}_{i2}^{\dag}\hat{\sigma}_{j2})\nonumber\\
&&+n_{1c}(\hat{\sigma}_{i2}^{\dag}\hat{\rho}\hat{\sigma}_{j2}+\hat{\sigma}_{j2}^{\dag}\hat{\rho}\hat{\sigma}_{i2}
       -\hat{\sigma}_{i2}\hat{\sigma}_{j2}^{\dag}\hat{\rho}-\hat{\rho}\hat{\sigma}_{j2}\hat{\sigma}_{i2}^{\dag})],\\
&\mathscr{L}_{2c}\hat{\rho}=&\sum^{2}_{i,j=1}\frac{\Gamma_{ijc}}{2}[(n_{2c}+1)(\hat{\sigma}_{gj}\hat{\rho}\hat{\sigma}_{gi}^{\dag}
+\hat{\sigma}_{gi}\hat{\rho}\hat{\sigma}_{gj}^{\dag}-\hat{\sigma}_{gj}^{\dag}\hat{\sigma}_{gi}\hat{\rho}-\hat{\rho}\hat{\sigma}_{gi}^{\dag}\hat{\sigma}_{gj})\nonumber\\
&&+n_{2c}(\hat{\sigma}_{gi}^{\dag}\hat{\rho}\hat{\sigma}_{gj}+\hat{\sigma}_{gj}^{\dag}\hat{\rho}\hat{\sigma}_{gi}
-\hat{\sigma}_{gi}\hat{\sigma}_{gj}^{\dag}\hat{\rho}-\hat{\rho}\hat{\sigma}_{gj}\hat{\sigma}_{gi}^{\dag})],\\
&\mathscr{L}_{3c}\hat{\rho}=&\frac{\gamma_{e}}{2}[(n_{e}+1)(2\hat{\sigma}_{21}\hat{\rho}\hat{\sigma}_{21}^{\dag}-\hat{\sigma}_{21}^{\dag}\hat{\sigma}_{21}
\hat{\rho}-\hat{\rho}\hat{\sigma}_{21}^{\dag}\hat{\sigma}_{21})\nonumber\\
&&+n_{e}(2\hat{\sigma}_{21}^{\dag}\hat{\rho}\hat{\sigma}_{21}-\hat{\sigma}_{21}\hat{\sigma}_{21}^{\dag}\hat{\rho}-\hat{\rho}\hat{\sigma}_{21}\hat{\sigma}_{21}^{\dag})],\\
&\mathscr{L}_{\Gamma}\hat{\rho}=&\sum^{2}_{i=1}\frac{\Gamma_{i}}{2}(2\hat{\sigma}_{\alpha ii}\hat{\rho}\hat{\sigma}_{\alpha ii}-\hat{\rho}\hat{\sigma}_{\alpha ii}-\hat{\sigma}_{\alpha ii}\hat{\rho}),\\
&\mathscr{L}_{\chi\Gamma}\hat{\rho}=&\sum^{2}_{i=1}\frac{\chi\Gamma_{i}}{2}(2\hat{\sigma}_{bi}\hat{\rho}\hat{\sigma}_{bi}^{\dag}-\hat{\rho}\hat{\sigma}
_{\alpha ii}-\hat{\sigma}_{\alpha ii}\hat{\rho})
\end{eqnarray}

Eq.(6) describes the effect of the high temperature reservoirs, where \(n_{h}\) denotes the average photon numbers of the high temperature reservoir, while the low temperature reservoir has the average phonon numbers \(n_{1c}=[exp(\frac{(E_{e_{2}}-E_{\alpha_{i}})}{k_{B}T_{a}})]^{-1}\) in Eq.(7), \(\gamma_{iic}=\gamma_{ic}(\gamma_{jjc}=\gamma_{jc})\) are the spontaneous decay rates from level \(|e_{2}\rangle\) to level \(|\alpha_{i}\rangle(i=1,2)\) respectively. The cross-coupling \(\gamma_{ijc}\) describes the effect of Fano interference. It is assumed that \(\gamma_{ijc}=\gamma_{jic}\) with \(\gamma_{ijc}=\eta_{1}\sqrt{\gamma_{ic}\gamma_{jc}}\) (i, j=1,2), where \(\eta_{1}\) describes the quantum interference intensity with \(\eta_{1}=1\) meaning the fully quantum interference and \(\eta_{1}=0\) for no interference. Similarly, another low temperature reservoir is described by Eq.(8) with \(n_{2c}=[exp(\frac{(E_{\beta_{i}}-E_{g})}{k_{B}T_{a}})]^{-1}\) being the cold reservoir phonon numbers. \(\Gamma_{ijc}=\Gamma_{jic}\) is defined by \(\Gamma_{ijc}=\eta_{2}\sqrt{\Gamma_{ic}\Gamma_{jc}}\) in Eq.(8), which describes the Fano interference induced by the spontaneous decay rates, \(\Gamma_{iic}\)=\(\Gamma_{ic}(\Gamma_{jjc}\)=\(\Gamma_{jc})\)(i, j=1,2) from level \(|\beta_{i,(i=1,2)}\rangle\) to level \(|g\rangle\) with \(\eta_{2}\) denoting the quantum interference robustness. In Eq.(9), \(n_{e}=[exp(\frac{(E_{e_{1}}-E_{e_{2}})}{k_{B}T_{a}})]^{-1}\) is the corresponding thermal occupation numbers of photon functioned by the robust coupled-dipoles J at temperature \(T_{a}\). \(\hat{\sigma}_{\alpha_{ii}}=|\alpha_{i}\rangle\langle\alpha_{i}|_{(i=1,2)}\) is defined in Eq.(10) and Eq.(11). Next, we will quantitatively depicted the electron transfer mechanisms in the photosynthetic RC via the density matrix dynamic element equations (seen in the Appendix).

\section{Results and discussion }

In this present model, we will explore the effect of the coupled-dipoles on the photosynthetic performance evaluated by the population dynamics of the two coupled sites, j-V characteristic and output power. For the sake of calculation, the output currents through two charge-separation pathways, \(|\alpha_{1}\rangle\) \(\rightarrow\) \(|\beta_{1}\rangle\) (Path 1) and \(|\alpha_{2}\rangle\) \(\rightarrow\) \(|\beta_{2}\rangle\) (Path 2) are set to be the same in the following discussion, i.e., the  rates of release  \(\Gamma_{1}\)=\(\Gamma_{2}\). Not only that, the total delivered voltage of the two channels is defined as the sum of their chemical potential differences between state \(|\alpha_{1}\rangle(|\alpha_{2}\rangle)\) and state \(|\beta_{1}\rangle(|\beta_{2}\rangle)\), \(eV\)=\(\sum\limits_{i=1}^{2}[ E_{\alpha_{i}}-E_{\beta_{i}}+k_{B}T_{a}\ln(\frac{\rho_{\alpha_{i}\alpha_{i}}}{\rho_{\beta_{i}\beta_{i}}})]\). The selected parameters (Shown in Table 1) \(E_{\alpha_{i}}\)and \(E_{\beta_{i}}\) \((i=1, 2)\) through the two charge-separation pathways are referred to Ref.\cite{CreatoreAn} with \(e\) being the elementary electron charge. Therefore, the performance of photosynthetic QHE in RC can be numerically evaluated by the current-voltage (\(j-V\)) characteristics and the generated power \(P=j\cdot V\).
Direct impact of the coupled-dipoles on the photosynthetic performance will be described by the population dynamics of the two coupled sites $|e_{1}\rangle$ and $|e_{2}\rangle$.

\begin{table}
\begin{center}
\caption{Model parameters used in the numerical calculations.}
\label{Table 1}
\vskip 0.3cm\setlength{\tabcolsep}{1cm}
\begin{tabular}{ccc}%{|c|c|c|c|c|c|c|c|c|c|}
\hline
\hline
                                                                        & Values                 & Units  \\
\hline
\(E_{D_{1}}-E_{g}=E_{D_{2}}-E_{g}\)                                     & 1.8                    & eV  \\
\(E_{D_{1}}-E_{\alpha_{1}}=E_{D_{1}}-E_{\alpha_{2}}\)                   & 0.2                    & eV  \\
\(E_{\beta_{1}}-E_{g}=E_{\beta_{2}}-E_{g}\)                             & 0.2                    & eV  \\
\(\gamma_{h}\)                                                          & \(2.48\times10^{-6}\)  & eV  \\
\(\gamma_{e}\)                                                          & 0.025                  & eV  \\
\(\gamma_{1c}=\gamma_{2c}\)                                             & 0.012                  & eV  \\
\(\Gamma_{1}=\Gamma_{2}\)                                               & 0.124                  & eV  \\
\(\Gamma_{1c}=\Gamma_{2c}\)                                             & 0.0248                 & eV  \\
\(n_{h}\)                                                               & 6000                   &     \\
\(T_{a}\)                                                               & 300                    &     \\
\(\chi\)                                                                & 0.2                    &     \\
\hline
\hline
\end{tabular}
\end{center}
\end{table}

\begin{figure}[htp]
\center
\includegraphics[width=0.35\columnwidth]{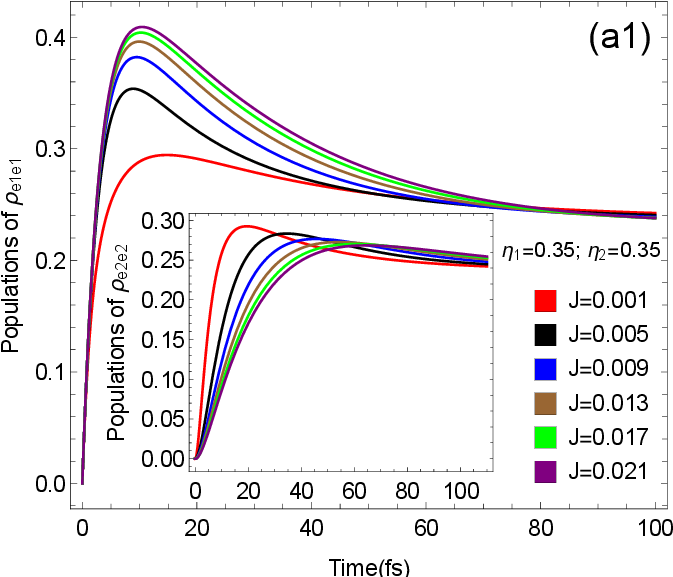 }\includegraphics[width=0.36\columnwidth]{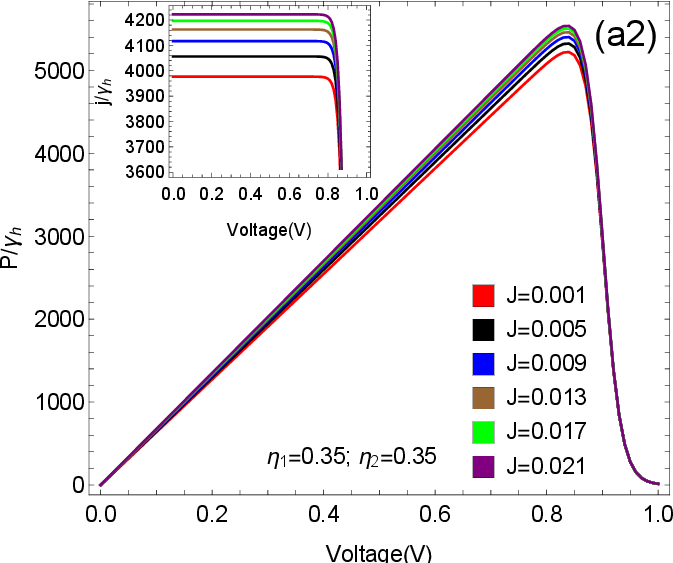 }
\caption{(Color online) (a1) Population dynamics of level $|e_{1}\rangle$ and level $|e_{2}\rangle$, (a2) j-V characteristic and power P of this QHE with different J =0.001, 0.005, 0.009, 0.013, 0.017 and 0.021. The quantum interference intensities are selected as $\eta_{1}$ =0.35 and $\eta_{2}$ =0.35. Other parameters are taken from Table \ref{Table 1}.}
\label{Fig.2}
\end{figure}

In Fig.\ref{Fig.2}, the population dynamics of states $|e_{1}\rangle$ and $|e_{2}\rangle$ are plotted in (a1) and its inset, and the numerical output power, steady-state j-V characteristic versus different robustness of the the coupled-dipoles are illustrated in (a2) and its inset with J=0.001, 0.005, 0.009, 0.013, 0.017 and 0.021. The quantum interference intensities are set as \(\eta_{1}=0.35\), and \(\eta_{2}=0.35\) in our numeral simulation. The antipodal influences on the two coupled sites $|e_{1}\rangle$ and  $|e_{2}\rangle$ are shown by the peak populations about in the range of [0, 100fs] in Fig.\ref{Fig.2} (a1). As for the excited state $|e_{1}\rangle$, the increasing population is generated by the more robust coupled-dipoles J. However, the peak populations of \(\rho_{e2e2}\) decrease with the increments of coupled-dipole interactions J in the same time range, [0, 100fs] in Fig.\ref{Fig.2} (a1). What's more, it notes that the influence J on the peak populations in the excited state $|e_{1}\rangle$ is sensitive, which can be drawn from the peak differences between the purple and red curves with J=0.001 and 0.021, respectively. The results indicate that more excited electrons are transferred to the exciton state $|e_{1}\rangle$ by the enhanced coupled-dipoles J. The large population differences indicate the better photosynthetic performance, which will be manifested by the j-V characteristic curves and output power in Fig.\ref{Fig.2} (a2). From the inset in Fig.\ref{Fig.2} (a2), the curves show that the increasing short-circuit currents and peak powers are achieved by the increments of J. The above results demonstrate the coupled-dipoles between the two different charge separation pathways throw a positive impact on its photosynthetic performance.

\begin{figure}[htp]
\center
\includegraphics[width=0.35\columnwidth]{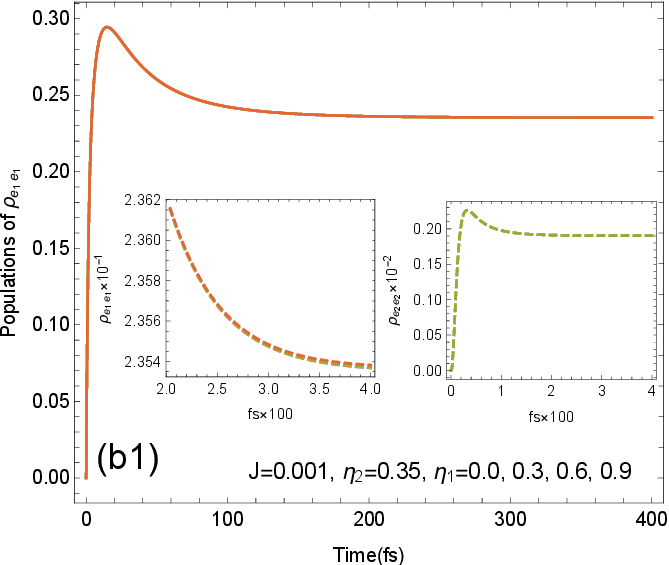 }\includegraphics[width=0.348\columnwidth]{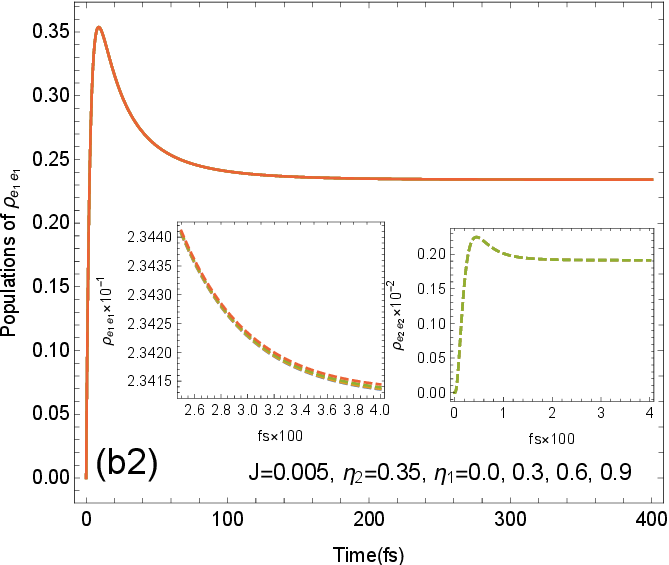 }{\hspace{0.2cm}}
\includegraphics[width=0.35\columnwidth]{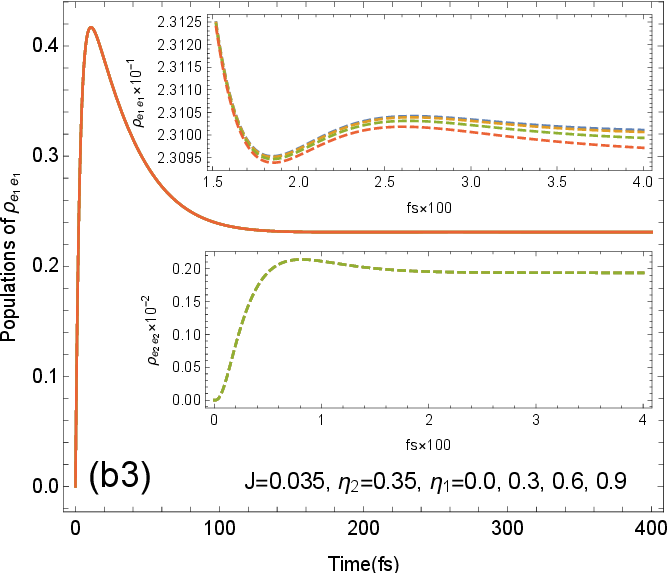 }\includegraphics[width=0.348\columnwidth]{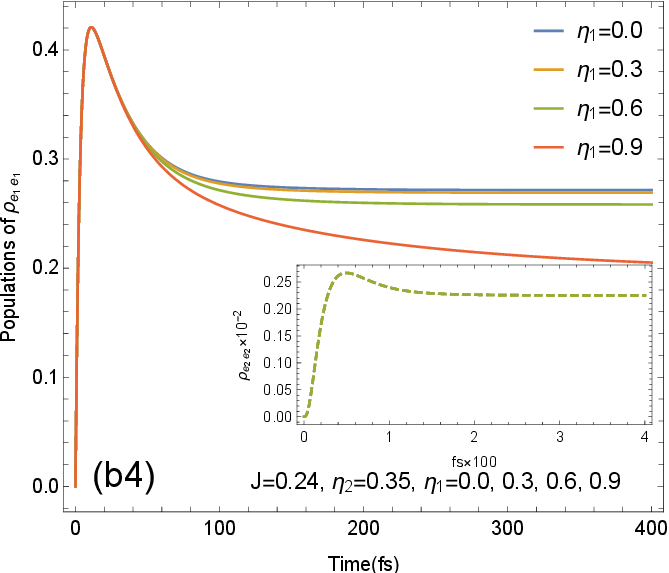 }
\caption{(Color online) Population dynamics of level $|e_{1}\rangle$ and level $|e_{2}\rangle$ under different quantum interference intensities $\eta_{1}$ =0, 0.3, 0.6, 0.9, and $\eta_{2}$ =0.35 with different coupled-dipoles robustness J=0.001, 0.005, 0.035 and 0.24 in (b1), (b2), (b3) and (b4) respectively. Other parameters are the same to those in Fig.\ref{Fig.2}.}
\label{Fig.3}
\end{figure}

In this proposed photosynthetic QHE model, what is the role of the coupled-dipoles between the two charge separation pathways in the Fano interference ? Fig.\ref{Fig.3} demonstrates that the Fano interference caused by the cross-coupling \(\gamma_{ijc}\) is deactivated owing to the coupled-dipoles. In Fig.\ref{Fig.3}, the population dynamics of states $|e_{1}\rangle$ and $|e_{2}\rangle$ are regulated by the Fano interference strength $\eta_{1}$. Counter-intuitively, we notice that the increasing Fano interference $\eta_{1}$ throws almost no influence on the population dynamics of state $|e_{2}\rangle$, and the zoomed insets of populated $|e_{1}\rangle$ show minimal influence when the coupled-dipoles J is in the range of 0.001 \(\leq\) J \(\leq\) 0.035 (cf. Fig.\ref{Fig.3} from (b1) to (b3) ), as is a similar result mentioned by the previous work\cite{Qin2016A}. When the coupled-dipoles J is in the range of 0.035 \(\leq\) J \(\leq\) 0.24, and now $\eta_{1}$ plays a bigger role after 60 fs (cf. the solid lines with different $\eta_{1}$ =0, 0.3, 0.6, 0.9), although its impact on their peaks of population of level $|e_{1}\rangle$ isn't significant.

Some previous work \cite{ZhangDelocalized} mentioned that certain ``dark" states can be generated by the dipole-dipole interactions in a photosynthetic process. Maybe we can take some cues for the physical phenomenon in this proposed model, and we speculate that the coupled-dipoles between the two charge separation pathways may also produces a dark state, which disables one chlorophylls located at the center of the PSII RC, forming one exciton state $|e_{1}\rangle$ being the primary electron donor. Thus, the Fano interference caused by the cross-coupling \(\gamma_{ijc}\) is deactivated. Changing the parameter $\eta_{1}$ has no substantial influence on the population dynamics of state  $|e_{2}\rangle$. But more electrons are transferred to the exciton state $|e_{1}\rangle$, which ultimately results in the increasing peak population on the exciton state $|e_{1}\rangle$.

\begin{figure}[htp]
\center
\includegraphics[width=0.35\columnwidth]{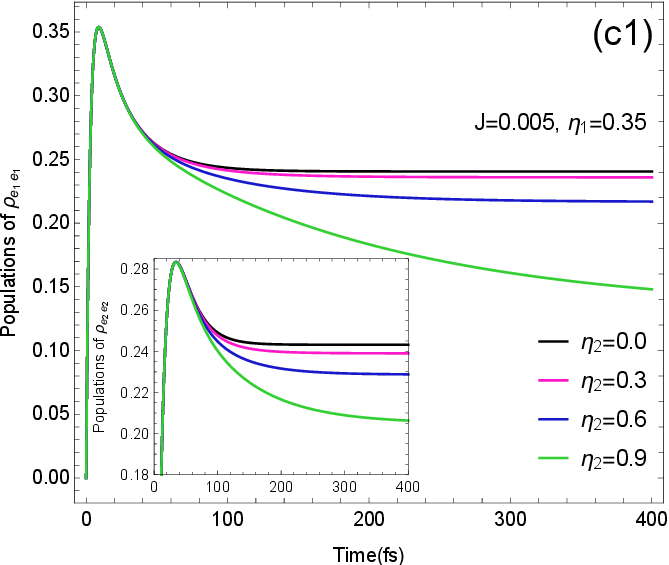 }\includegraphics[width=0.35\columnwidth]{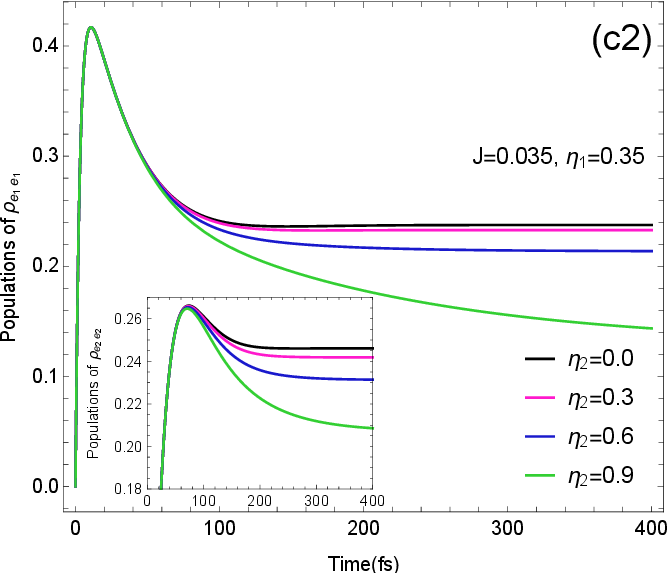 }
\includegraphics[width=0.35\columnwidth]{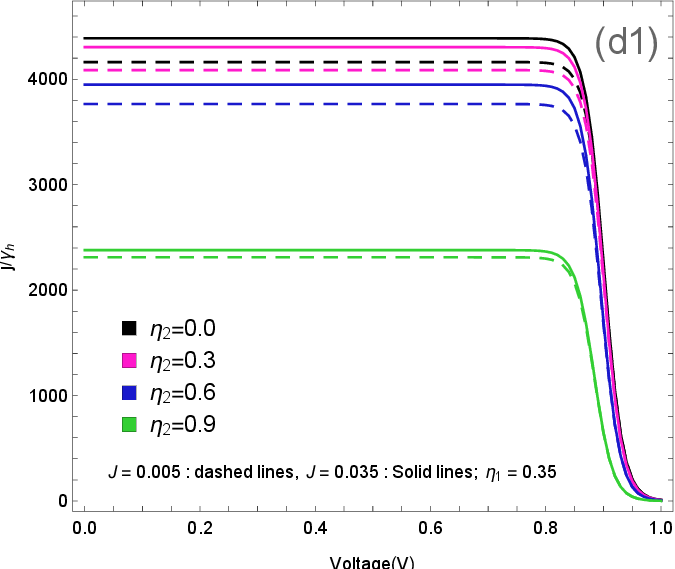 }\includegraphics[width=0.35\columnwidth]{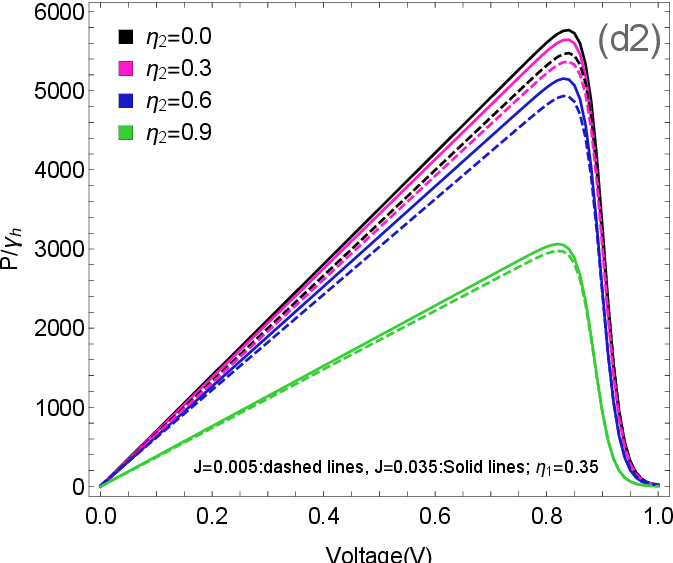 }
\caption{(Color online) Population dynamics of level $|e_{1}\rangle$ and level $|e_{2}\rangle$ in (c1) and (c2), and j-V characteristic and power P in (d1) and (d2) with different quantum interference intensities $\eta_{2}$ =0, 0.3, 0.6, 0.9, and $\eta_{1}$ =0.35 under different robustness of the coupled-dipoles J=0.005 and J=0.035.  Other parameters are the same to those in Fig.\ref{Fig.2}}
\label{Fig.4}
\end{figure}

In this photosynthetic QHE system, the influence of J on another quantum interference originating from the crossing spontaneous decay rates, \(\Gamma_{1c}\) and \(\Gamma_{2c}\) from level \(|\beta_{i,(i=1,2)}\rangle\) to level \(|g\rangle\) stimulates our curiousness. Therefore, we similarly focus on the influence involving the robustness of J on the quantum interference described by $\eta_{2}$. It notes that the inactive short-lived quantum interference $\eta_{2}$ still appears about in the range of [0, 100fs].  With the time evolution, the influence of $\eta_{2}$ gradually works on the population dynamics of levels $|e_{1}\rangle$ and $|e_{2}\rangle$ in the subsequent time range shown in Fig.\ref{Fig.4} (c1) and (c2). We also notice that the increasing robustness of J brings about an increment in the population difference between $|e_{1}\rangle$ and $|e_{2}\rangle$, which will indicate an enhancement of photosynthetic performance in this QHE model. The j-V characteristic and power P in (d1) and (d2) demonstrate this prediction. Curves in Fig.\ref{Fig.4} (d1) and (d2) with the same color signify the identical quantum interference $\eta_{2}$. It notes that the j-V characteristic and power P increase with the coupled-dipoles robustness J but decrease with $\eta_{2}$. And the change from the dashed to solid lines with the same color manifests the enhanced photosynthetic performance due to the increasing coupled-dipoles robustness J. In this photosynthetic QHE system, the increasing $\eta_{2}$ indicates more electrons transiting to the state \(|g\rangle\), which results in the decreasing j-V characteristic and power P in Fig.\ref{Fig.4} (d1) and (d2). The curves with the same color illustrate that the negative function of quantum interference $\eta_{2}$ can be suppressed by the coupled-dipoles robustness J. Finally we would like to point out that these conclusions are drawn from a typical photosynthetic RC model, it may have universal theoretical significance for the design of artificial photosynthetic devices.

\section{Conclusion}

In this proposed photosynthetic QHE model, the roles of the coupled-dipoles between two-charge separation pathways were quantitatively explored in the photosynthetic performance. The results evaluated by the population dynamics, j-V characteristic and power P demonstrated that the photosynthetic performance was greatly enhanced but quantum interference was deactivated by the coupled-dipoles, i.e., the increasing J makes $\eta_{1}$ have minimal effect on the populated $|e_{1}\rangle$ but no effect on the populated $|e_{2}\rangle$ which brings out more electrons transferred to the exciton state $|e_{1}\rangle$ and ultimately results in the increasing peak population on the exciton state $|e_{1}\rangle$. However, the increasing J is beneficial to the enhancement of the photovoltaic performance evaluated by j-V characteristic and power P, although the increasing J makes $\eta_{2}$ have a negative influence. We argue that it is a new operating regime of the coupled-dipoles between two-charge separation pathways in the energy transport process in the biological photosynthetic, and some artificial strategies may be motivated by this photosynthetic QHE model.

\section{Acknowledgments}

S. C. Zhao is grateful for funding from the National Natural Science Foundation of China (grants 62065009 and  61565008) and General Program of Yunnan Applied Basic Research Project, China (grant 2016FB009).

\section{APPENDIX}

The density matrix dynamic element equations are given as follows:

\begin{eqnarray}
&\dot{\rho}_{e_{1}e_{1}}=&-\gamma_{e}[(n_{e}+1)\rho_{e_{1}e_{1}}-n_{e}\rho_{e_{2}e_{2}}]-\gamma_{h}[(n_{h}+1)\rho_{e_{1}e_{1}}-n_{h}\rho_{gg}],\nonumber\\
&\dot{\rho}_{e_{2}e_{2}}=&\gamma_{e}[(n_{e}+1)\rho_{e_{1}e_{1}}-n_{e}\rho_{e_{2}e_{2}}]-\gamma_{1c}[(n_{1c}+1)\rho_{e_{2}e_{2}}-n_{c1}\rho_{\alpha_{1}\alpha_{1}}]\nonumber\\
                         &&-\gamma_{2c}[(n_{1c}+1)\rho_{e_{2}e_{2}}-n_{1c}\rho_{\alpha_{2}\alpha_{2}}]+2\gamma_{12c}n_{c1}Re[\rho_{\alpha_{1}\alpha_{2}}],\nonumber\\
&\dot{\rho}_{\alpha_{1}\alpha_{1}}=&\gamma_{1c}[(n_{1c}+1)\rho_{e_{2}e_{2}}-n_{1c}\rho_{\alpha_{1}\alpha_{1}}]-\gamma_{12c}n_{1c}Re[\rho_{\alpha_{1}\alpha_{2}}]\nonumber\\
                                   &&-(1+\lambda)\Gamma_{1}\rho_{\alpha_{1}\alpha_{1}},\nonumber\\
&\dot{\rho}_{\alpha_{2}\alpha_{2}}=&\gamma_{1c}[(n_{1c}+1)\rho_{e_{2}e_{2}}-n_{1c}\rho_{\alpha_{1}\alpha_{1}}]-\gamma_{12c}n_{1c}Re[\rho_{\alpha_{1}\alpha_{2}}]\nonumber\\
                                   &&-(1+\lambda)\Gamma_{2}\rho_{\alpha_{2}\alpha_{2}},\nonumber\\
&\dot{\rho}_{\alpha_{1}\alpha_{2}}=&-i\triangle_{1}\rho_{\alpha_{1}\alpha_{2}}-\frac{1}{2}(\gamma_{1c}+\gamma_{2c})n_{1c}\rho_{\alpha_{1}\alpha_{2}}\nonumber\\
                                   &&+\frac{1}{2}\gamma_{12c}[2(n_{1c}+1)\rho_{e_{2}e_{2}}-n_{1c}\rho_{\alpha_{2}\alpha_{2}}-n_{1c}\rho_{\alpha_{1}\alpha_{1}}]\nonumber\\
&\dot{\rho}_{\beta_{1}\beta_{1}}=&\Gamma_{1}\rho_{\alpha_{1}\alpha_{1}}-\Gamma_{1c}[(n_{2c}+1)\rho_{\beta_{1}\beta_{1}}-n_{2c}\rho_{gg}]
                                 -\Gamma_{12c}(n_{2c}+1)Re[\rho_{\beta_{1}\beta_{2}}],\nonumber\\
&\dot{\rho}_{\beta_{2}\beta_{2}}=&\Gamma_{2}\rho_{\alpha_{2}\alpha_{2}}-\Gamma_{2c}[(n_{2c}+1)\rho_{\beta_{2}\beta_{2}}-n_{2c}\rho_{gg}]
                                 -\Gamma_{12c}(n_{2c}+1)Re[\rho_{\beta_{1}\beta_{2}}],\nonumber\\
&\dot{\rho}_{\beta_{1}\beta_{2}}=&-i\triangle_{2}\rho_{\beta_{1}\beta_{2}}-\frac{1}{2}(\Gamma_{1c}+\Gamma_{2c})(n_{2c}+1)\rho_{\beta_{1}\beta_{2}}\nonumber\\
                                 &&-\frac{1}{2}\Gamma_{12c}[(n_{2c}+1)\rho_{\beta_{1}\beta_{1}}+(n_{2c}+1)\rho_{\beta_{2}\beta_{2}}
                                 -2n_{2c}\rho_{gg}],\nonumber\\
&\rho_{gg}=&1-\rho_{e_{1}e_{1}}-\rho_{e_{2}e_{2}}-\rho_{\alpha_{1}\alpha_{1}}-\rho_{\alpha_{2}\alpha_{2}}-\rho_{\beta_{1}\beta_{1}}-\rho_{\beta_{2}\beta_{2}}.
\end{eqnarray}

\noindent where \(\triangle_{1}=E_{\alpha_{1}}-E_{\alpha_{2}}\) and \(\triangle_{2}=E_{\beta_{1}}-E_{\beta_{2}}\) are the splitting of the states \(|\alpha_{1}\rangle(|\alpha_{2}\rangle)\) and \(|\beta_{1}\rangle(|\beta_{2}\rangle)\). We utilize the equations to simulate dynamics of the reaction centre.

\bibliography{reference}
\bibliographystyle{unsrt}%abbrv, acm, alpha, apalike, ieeetr, plain, siam , unsrt
%\end{CJK*}  %% end the Chinese environment
\end{document}